\documentclass[aps,prd,twocolumn,a4paper,superscriptaddress,floatfix]{revtex4-1}
\usepackage{graphicx}
\usepackage{hyperref}
\usepackage{amssymb}
\usepackage{natbib}
\usepackage{color}
\usepackage[dvipsnames]{xcolor}
\usepackage{amsmath}
\begin{document}
\def\cjm#1{{\color{ForestGreen} \sf #1}}
\newcommand{\be}{\begin{equation}}
\newcommand{\ee}{\end{equation}}
\newcommand{\bq}{\begin{eqnarray}}
\newcommand{\eq}{\end{eqnarray}}
\newcommand{\bw}{\begin{widetext}}
\newcommand{\ew}{\end{widetext}}
\newcommand{\bsq}{\begin{subequations}}
\newcommand{\esq}{\end{subequations}}
\newcommand{\bc}{\begin{center}}
\newcommand{\ec}{\end{center}}
\title{Quantitative constraints on modified gravity paradigms}
\author{S. R. Pinto}
\email{up202004386@edu.fc.up.pt}
\affiliation{Centro de Astrof\'{\i}sica da Universidade do Porto, Rua das Estrelas, 4150-762 Porto, Portugal}
\affiliation{Faculdade de Ci\^encias, Universidade do Porto, Rua do Campo Alegre, 4150-007 Porto, Portugal}
\author{A. M. Cabral}
\email{up201906538@edu.fc.up.pt}
\affiliation{Centro de Astrof\'{\i}sica da Universidade do Porto, Rua das Estrelas, 4150-762 Porto, Portugal}
\affiliation{Faculdade de Ci\^encias, Universidade do Porto, Rua do Campo Alegre, 4150-007 Porto, Portugal}
\author{C. J. A. P. Martins}
\email{Carlos.Martins@astro.up.pt}
\affiliation{Centro de Astrof\'{\i}sica da Universidade do Porto, Rua das Estrelas, 4150-762 Porto, Portugal}
\affiliation{Instituto de Astrof\'{\i}sica e Ci\^encias do Espa\c co, CAUP, Rua das Estrelas, 4150-762 Porto, Portugal}

\begin{abstract}
We use low-redshift background cosmology data to place quantitative constraints on three separate modified gravity models, each of which aims to explain the low-redshift acceleration through a different physical mechanism. The Lifshitz cosmology is effectively a parametric extension of the canonical $\Lambda$CDM model, where a time-dependent cosmological constant originates from vacuum energy. The Infinite Statistics model is also a parametric extension of $\Lambda$CDM, where the dark energy is dynamic and originates from the curvature of a dual space-time. We show that the data restricts the additional parameters in these models to be consistent with their $\Lambda$CDM values, and in particular that it implies that the theoretically predicted value for a dimensionless coupling parameter in the Lifshitz model is ruled out at more than six standard deviations. In the Regge-Teitelboim model, gravity is described by embedding the usual space-time manifold in a fixed higher-dimensional background, and there is no parametric $\Lambda$CDM limit. We study several separate realizations of the model, respectively introduced by Davidson, by Fabi \textit{et al.}, and by Stern \& Xu, and show that the first two are ruled out by the low-redshift data we use, while the latter is consistent with this data but requires a non-standard value of the matter density. Overall, our analysis highlights the tight constraints imposed by current data on the allowed low-redshift deviations from the standard $\Lambda$CDM background evolution.
\end{abstract}
\date{\today}
\maketitle

\section{Introduction}

The most compelling goal of modern fundamental cosmology is identifying the physical mechanism underlying the observed low-redshift acceleration of the universe. Conceptually, there are three theoretical possibilities that can be envisaged. The first possibility is a cosmological constant, which has the minimal number of additional model parameters. This solution is, broadly speaking, in agreement with the currently available data, even though the observational inferred value is entirely unexpected given contemporary theoretical expectations. The second possibility is adding dynamical degrees of freedom, particularly in the form of scalar fields, since they are known to be among Nature's building blocks. Finally, the third and most radical possibility is modifying the behaviour of gravity.

In this work we present a comparative study of the observational constraints on three different modified gravity models. All three have been suggested, in the recent literature, as alternatives to the canonical $\Lambda$CDM paradigm, but these claims were based on somewhat qualitative arguments. Here we compare the models with low-redshift background cosmology data, obtaining statistically robust constraints on each of them. Each of the models stems from different postulates or physical assumptions. The Lifshitz model, recently explored by Berechya \& Leonhardt \cite{Berechya}, proposes that dark energy originates from vacuum fluctuations, with the acceleration of the Universe coming from treating it as a time-dependent dielectric medium. The infinite statistics model, introduced by Jejjala \textit{et al.} \cite{Jejjala1,Jejjala2} suggests that an effectively time-dependent vacuum energy density is a consequence of the geometry of the dual space time. Finally, the Regge-Teitelboim model, some of whose cosmological solutions have been considered by several authors \cite{Davidson,Fabi,Stern} is based on embeddings of our the space-time manifold in a fixed higher-dimensional background, and regards the embedding coordinates as dynamical degrees of freedom. These embeddings lead to additional source terms in the Einstein equations which, in certain circumstances, could lead to accelerating solutions.

Our analysis takes all three models at face value and phenomenologically constrains them using low-redshift background cosmology data. Perturbation theory methodology has not yet been applied to these models but, as we demonstrate, even background cosmology data is sufficient to provide highly stringent constraints. In the next session we briefly summarize the Type Ia supernova and Hubble parameter data and methods used in our analysis, both of which are standard.

In the following three sections we describe the motivation for each of the models (as suggested by the authors of each of them) and the present our constraints on the model parameters, obtained from the aforementioned data sets though standard likelihood analyses. Since we are concerned with the low-redshift behaviour of these models and also only using low-redshift data, we will ignore the contribution of radiation to each model's Friedmann equation. Finally in Sect. \ref{concl} we present a comparative analysis of our results as well as some conclusions.

\section{Data and methods}

Our analysis relies on two recent and independent datasets, each of which has been extensively used in the literature for such analyses. The first is the Pantheon catalogue \cite{Scolnic,Riess}, including its covariance matrix. The second is a compilation of 38 Hubble parameter measurements reported in Farooq \textit{et al.} \cite{Farooq}. Overall, all data is at redshifts $z<2.5$, and therefore the assumption of ignoring the radiation density has no significant impact in our results.

We follow a standard likelihood analysis---see, e.g. \cite{Verde}--- with the likelihood being defined as
\be
{\cal L}(q)\propto\exp{\left(-\frac{1}{2}\chi^2(q)\right)}\,,
\ee
where $q$ symbolically denotes the free parameters in the model being considered. Since our two data sets are independent, the total chi-square is the sum of the two, $\chi^2=\chi^2_{SN}+\chi^2_{HZ}$. We will generally work with the dimensionless Friedmann equation, and our observable for both datasets will be the re-scaled Hubble parameter, \be
E(z)=\frac{H(z)}{H_0}\,.
\ee
The canonical confidence levels are then identified, in terms of the corresponding $\Delta\chi^2$, with standard numerical tools.

The Pantheon supernova  dataset contains 1048 supernovas, which span the redshift range $0<z<2.3$ \cite{Scolnic}, further compressed into 6 correlated measurements of $E^{-1}(z)$ in the redshift range $0.07<z<1.5$ \cite{Riess}. The latter work demonstrates that this provides an effectively identical characterization of the dark energy properties as the full supernova sample, while the data compression has obvious computational advantages. The chi-square in this case has the explicit form
\be
\chi^2_{SN}(q)=\sum_{i,j}\left(E_{obs,i}-E_{mod,i}(q)\right)C_{ij}^{-1}\left(E_{obs,j}-E_{mod,j}(q)\right)\,,
\ee
where the obs and mod subscripts denote observations and model respectively, and $C$ is the covariance matrix of the dataset. We note that this analysis is independent of the Hubble constant, $H_0$.

The Farooq {\it et al.} Hubble parameter dataset is a heterogeneous set of 38 measurements some of which come from cosmic chronometers and the rest from baryon acoustic oscillations (BAO). We must point out that strictly speaking the BAO measurements rely on some underlying assumptions on a fiducial model. That having been said, this model dependence is known not be significant, at least for models close to $\Lambda$CDM, in which case this dependence is at the percent level---discussions can be found in \cite{BAO1,BAO2}. Previous works indicate that such model dependence can be larger in models with late-time inhomogeneities or those where backreaction is important. There is no particular reason why this should be the case for the models which we study in the following sections, but nevertheless this could be seen as a potential theoretical systematic. The correlation between these BAO measurements is non-zero but small, so the measurements in the dataset can be assumed to be independent (i.e., the covariance matrix is assumed to be trivial). The cosmic chronometer subset is less constraining than its BAO counterpart. In fairness, we should also point out that there are potential observational systematics associated with these measurements \cite{Concas,Vazdekis}.

In order to do the analysis in terms of $E(z)$ and thus esaily combine the two datasets in the previously defined likelihood we must marginalize the Hubble constant. This can be done analytically \cite{Anagnostopoulos}, thereby reducing the parameter space and also eliminating the need of choosing specific priors on the Hubble constant. The marginalization procedure relies on computing three separate quantities
\bq
A(q)&=&\sum_{i}\frac{E_{model,i}^2(q)}{\sigma^2_i}\\
B(q)&=&\sum_{i}\frac{E_{model,i}(q) H_{obs,i}}{\sigma^2_i}\\
C(q)&=&\sum_{i}\frac{H_{obs,i}^2}{\sigma^2_i}
\eq
where the $\sigma_i$ are the uncertainties in observed values of the Hubble parameter. Then the chi-square is given by
\be
\chi^2(q)=C(q)-\frac{B^2(q)}{A(q)}+\ln{A(q)}-2\ln{\left[1+Erf{\left(\frac{B(q)}{\sqrt{2A(q)}}\right)}\right]}
\ee
where $Erf$ is the Gauss error function and $\ln$ is the natural logarithm.

\section{The Lifshitz model}

Lifshitz cosmology stems from the hypothesis that curved spacetime is analogous to a dielectric medium whose index of refraction changes proportionally to the scale factor \cite{Leonhardt}. If that is the case, then the universe's vacuum energy can be calculated using methods developed long ago by Lifshitz and others \cite{Lifshitz}, and the dark energy could be due to these vacuum fluctuations. Ignoring the density parameter for radiation, the self–consistent dynamics that results from the interaction between the vacuum energy and the background universe can be written
\begin{equation}
    H^2(z)=H_0^2\left[\Omega_m(1+z)^{3}+\Omega_{LC}(z)\right]
    \label{eq:8}
\end{equation}
\begin{equation}
    H^2_0\dot{\Omega}_{LC}=8\alpha_{\Lambda}H\partial^3_tH^{-1}\,,
    \label{eq:9}
\end{equation}
where $\Omega_m=k \rho_0/3H^2_0$ is the usual density parameter for matter, $\Omega_{LC}$ is the (time/redshift dependent) density parameter for dark energy and $\alpha_{\Lambda}$ is a dimensionless coupling parameter which depends on the chosen cut–off and on the possible contributions of other fields in the standard model of particle physics. Assuming that the electromagnetic field provides the dominant contribution (an assumption which underlies the dielectric medium analogy) and a Planck-scale cutoff, its theoretically predicted value is \cite{Leonhardt}
\begin{equation}
    \alpha_\Lambda=\frac{1}{9\pi}\sim0.0354\,.
\end{equation}

Assuming that the contribution of the vacuum energy is negligible at last–scattering, the analysis of Berechya \& Leonhardt \cite{Berechya} presents an approximate closed-form solution of the Friedmann equation, valid at low redshifts
\begin{widetext}
\be
E^2(z)=\Omega_m(1+z)^3+\Omega_\infty\left[1+18\alpha_\Lambda\left(\ln{\left(1+\frac{\Omega_m}{\Omega_\infty}(1+z)^3\right)}-3\frac{\Omega_m(1+z)^3}{\Omega_m(1+z)^3+\Omega_\infty}\right)\right]\,.
\ee
\end{widetext}
where $\Omega_{\infty} \equiv \lim_{a \to\infty}\Omega_{LC}$ is an integration constant which one might physically interpret as the value of the quantum vacuum contribution to dark energy in the asymptotic future.

\begin{figure}
\begin{center}
\includegraphics[width=\columnwidth]{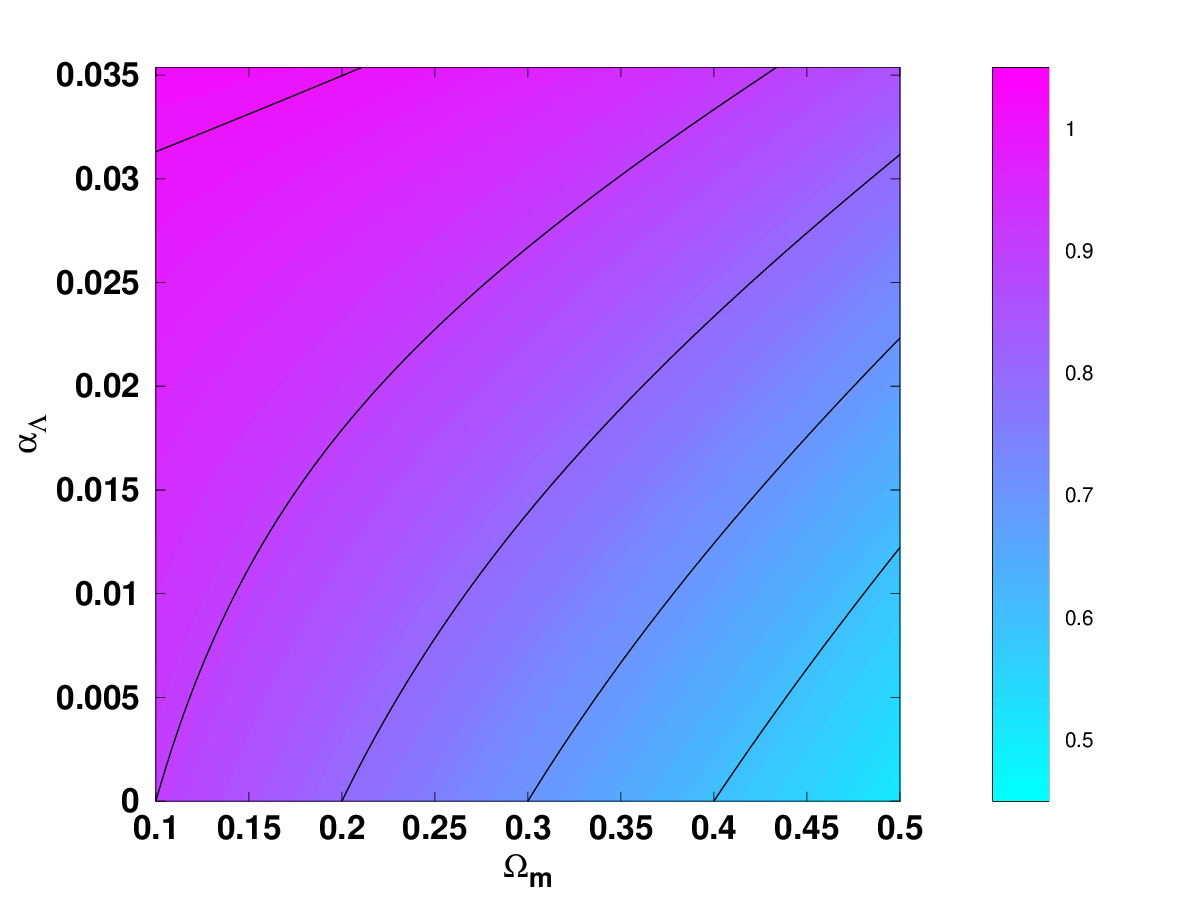}
\caption{Result of the numerical solution of Eq. \ref{consistencyL} in the relevant range of the model parameters. The color map represents the value of $\Omega_\infty$. The black solid lines identify the locus of values of $\Omega_\infty$ of 0.6, 0.7, 0.8, 0.9 and 1.0, respectively from bottom right to top left.}\label{fig1}
\end{center}
\end{figure}

Importantly, there is a normalization condition, since by definition we must have $E(0)=1$. This leads to 
\be\label{consistencyL}
\Omega_m+\Omega_\infty\left[1+18\alpha_\Lambda\left(\ln{\left(1+\frac{\Omega_m}{\Omega_\infty}\right)}-3\frac{\Omega_m}{\Omega_m+\Omega_\infty}\right)\right]=1\,.
\ee
This is clearly well-behaved in the $\Lambda$CDM limit $\alpha_\Lambda\to0$. The $\Omega_\infty\to0$ limit would correspond to an Einstein-de Sitter universe, with $\Omega_m=1$. Taking these at face value, the model is therefore a one-parameter extension of the flat $\Lambda$CDM model. The fact that $E(z)$ is a low-redshift analytic approximation is manifest in the fact that for large enough values of $\alpha_\Lambda$ the previous consistency condition can lead to $\Omega_\infty>1$. Figure \ref{fig1} depicts the numerical solution of Eq. \ref{consistencyL} in the relevant range of the model parameters, specifically $\alpha_\Lambda\in[0,1/9\pi]$ and $\Omega_m\in[0.1,0.5]$. In particular, in the limit $\Omega_m\to0$ we find
\be
\Omega_\infty=1-\Omega_m(1-36\alpha_\Lambda)\,;
\ee
for the theoretically expected value $\alpha_\Lambda\sim0.0354$ we therefore obtain $\Omega_\infty\simeq1+0.273\Omega_m$.

\begin{figure}
\begin{center}
\includegraphics[width=\columnwidth]{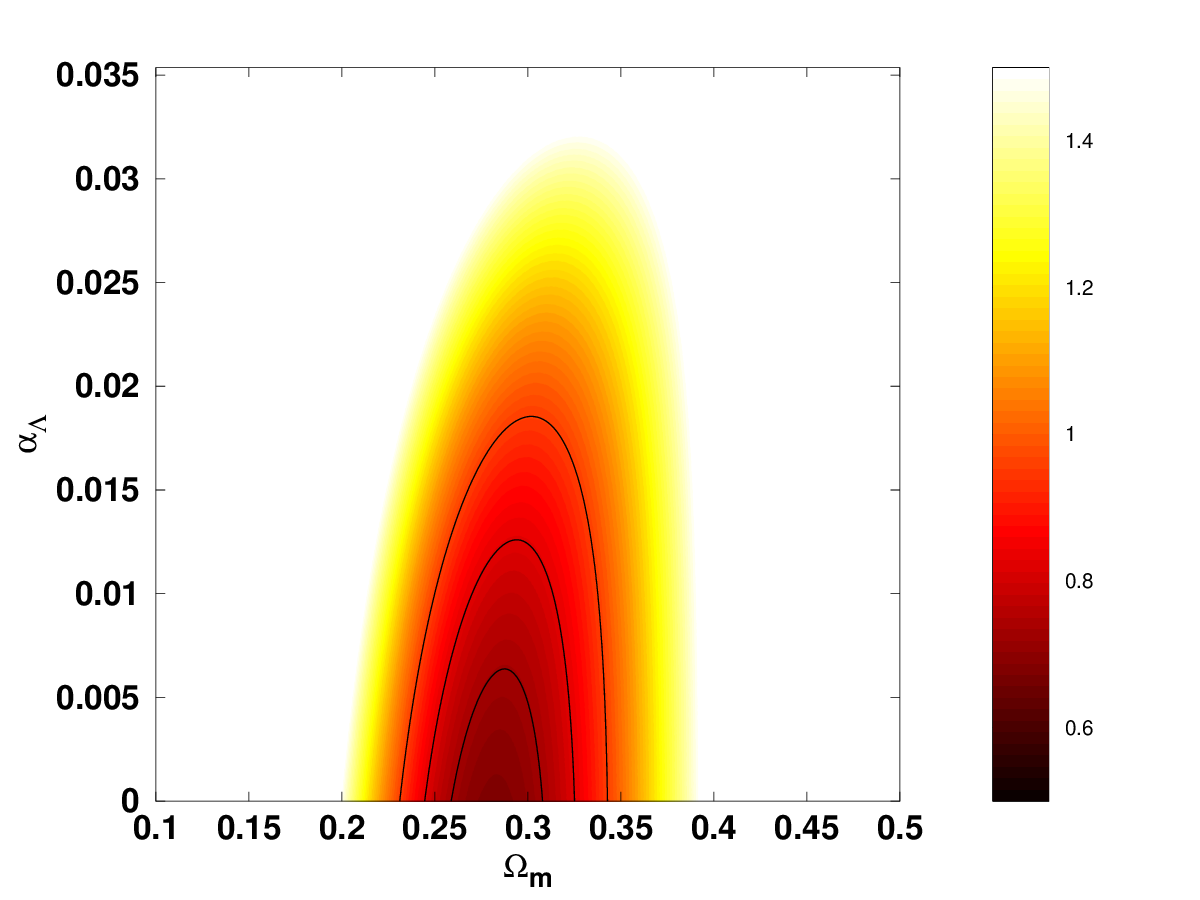}
\includegraphics[width=\columnwidth]{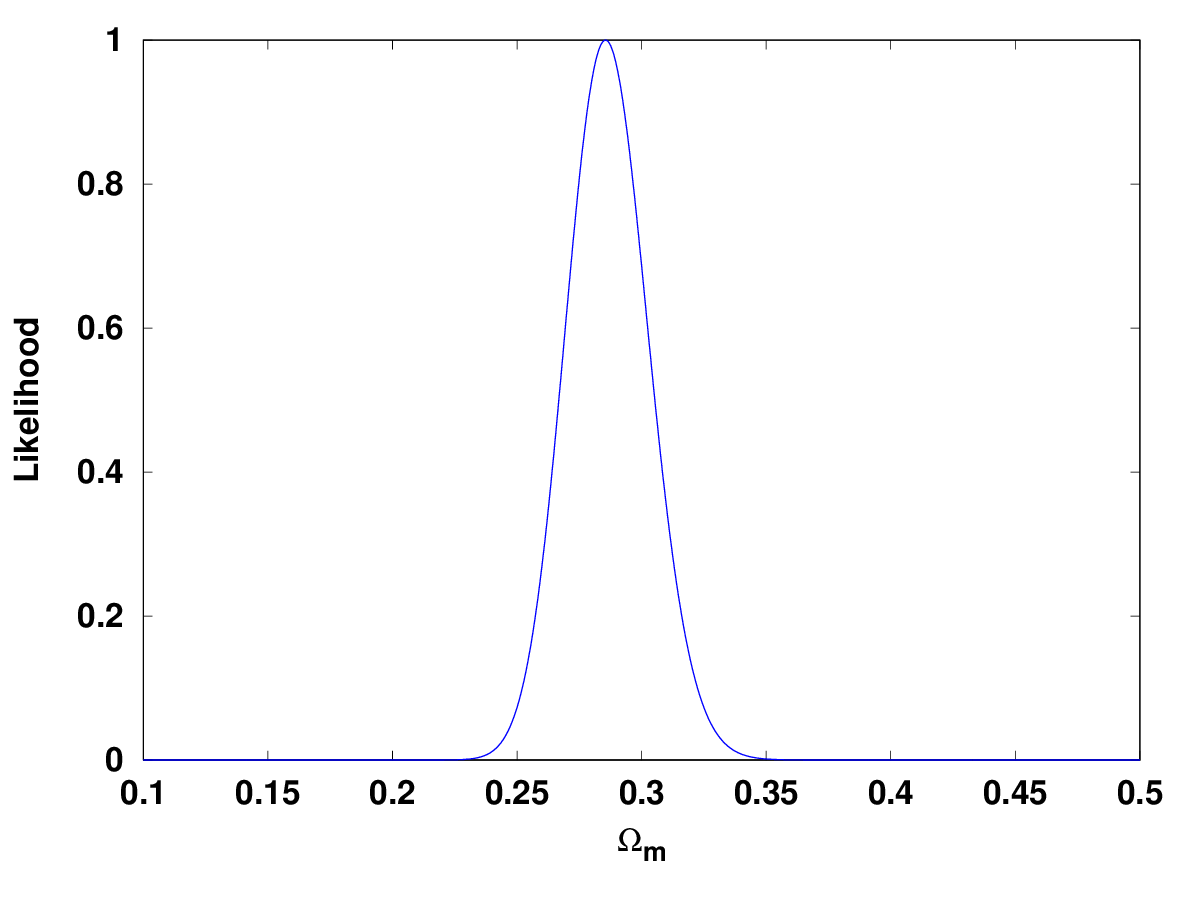}
\includegraphics[width=\columnwidth]{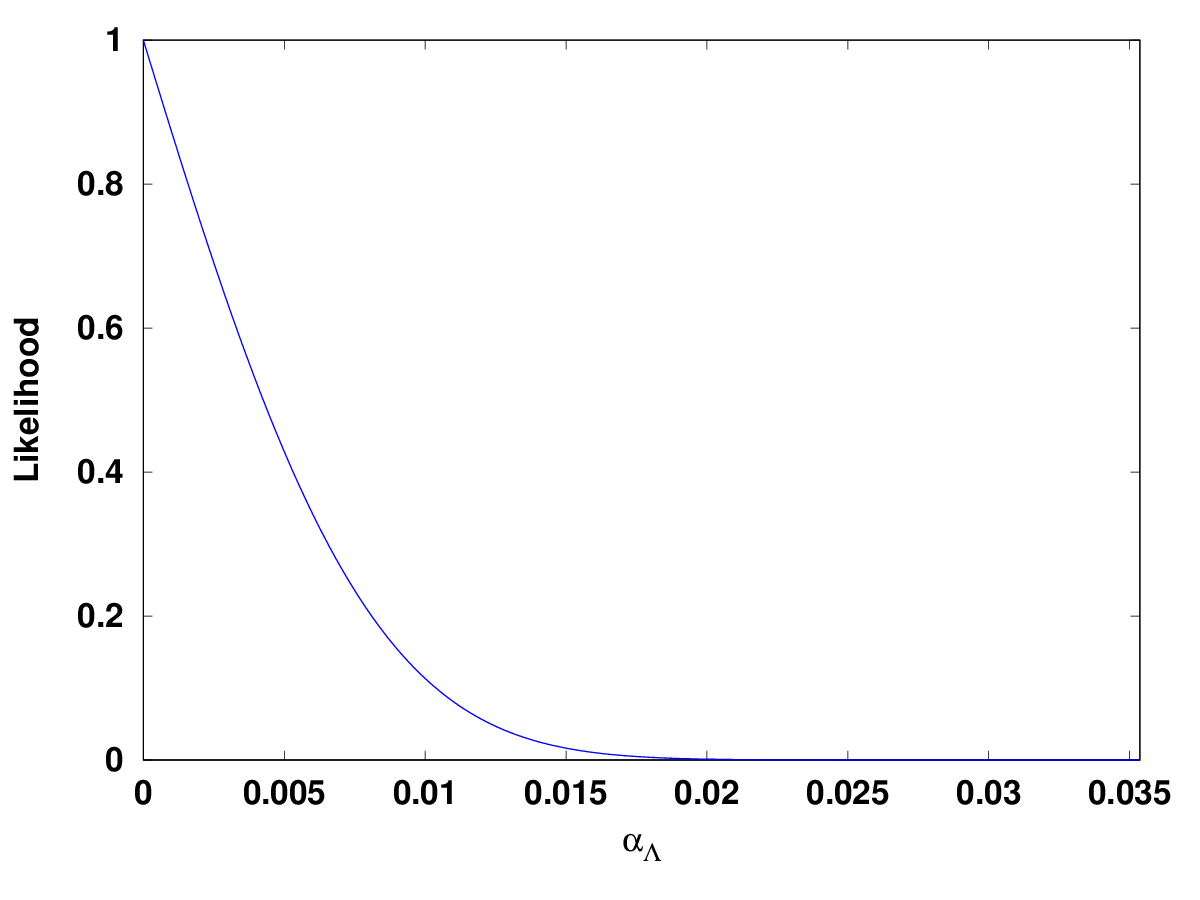}
\caption{Constraints on the Lifshift model, under the assumptions specified in the main text. In the top panel the black lines shows the one, two and three sigma confidence levels, while the color map depicts the reduced chi-square. The middle and bottom panels show the one-dimensional posterior likelihoods for each parameter. Note that the theoretically preferred value of $\alpha_\Lambda=1/9\pi$ corresponds to the top edge of the top panel and the right-side edge of the bottom panel.}\label{fig2}
\end{center}
\end{figure}

The work of \cite{Berechya} suggests, based on a qualitative analysis of particular choices of model parameters, that the model can reproduce the observed low-redshift acceleration of the universe. However, there is no thorough exploration of the model's parameter space, carried out with a statistically robust analysis, which we present in what follows. The results of this analysis, assuming uniform priors for the two model parameters with the ranges specified in the previous paragraph, are shown in Figure \ref{fig2}. It is noteworthy that there is no significant degeneracy between the two model parameters. The one-sigma constraint on the matter density is
\begin{equation}
    \Omega_{m}= 0.29 \pm 0.02\,,
\end{equation}
while for the dimensionless coupling we obtain the two-sigma upper limit
\begin{equation}
    \alpha_{\Lambda}<0.0094\,.
\end{equation}
This is fully consistent with the $\Lambda$CDM model, and in particular the theoretically preferred value of $\alpha_\Lambda=1/9\pi$ is excluded at about 6.5 standard deviations.

\section{The infinite statistics model}

The so-called infinite statistics model is based on the hypothesis that dark energy originates from the curvature of an assumed dual space \cite{Jejjala1}, which according to these authors can be motivated from quantum gravity, under the further assumptions of non-locality of the effective spacetime description and of Lorentz covariance. In that case the functional form of the dark energy contribution, viz. its redshift dependence, would be obtainable from the infinite statistics \cite{Infinite1,Infinite2} (technically defined as the unique statistics consistent with Lorentz covariance in the presence of non-locality) of the quanta of said dual space-time. In that case dark energy contributions would come from a series of terms: the lowest order of these would be a standard cosmological constant ($\Lambda$ that is realized as a dynamical geometry of the dual space-time), but there will be additional redshift-dependent terms, leading to modified Einstein equations.

\begin{figure}
\begin{center}
\includegraphics[width=\columnwidth]{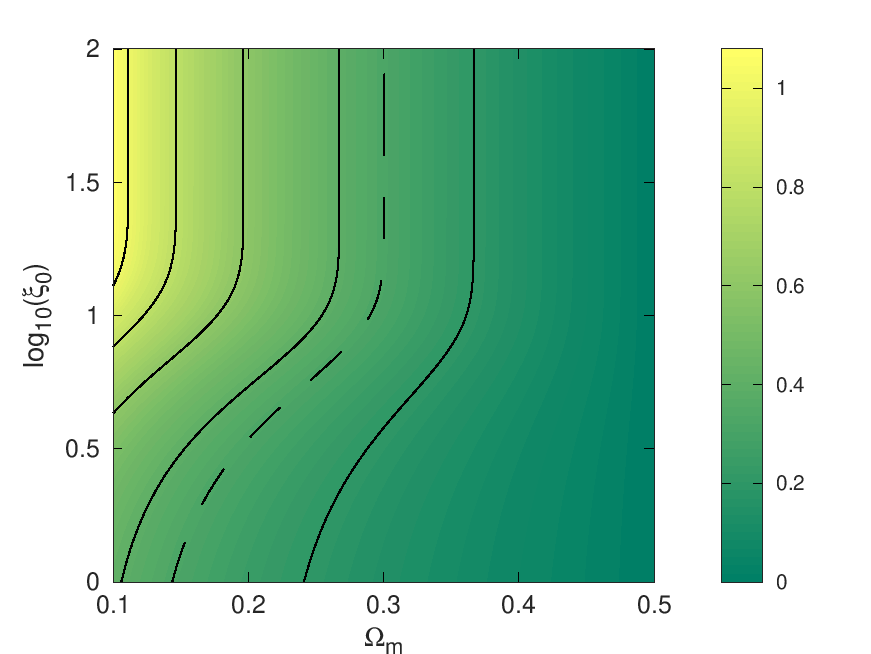}
\caption{The redshift at which the two terms in the Friedmann equation have identical values, as a function of the matter density and the scaling parameter, depicted by the color map. The black solid lines identify the locus of values of this redshift of 0.2, 0.4, 0.6, 0.8 and 1.0, respectively from the right to the left of the plot. The value of this redshift for a flat $\Lambda$CDM model is $z\sim0.326$, and is identified by the dashed line.}\label{fig3}
\end{center}
\end{figure}

Skipping the technical details which can be found in Jejjala \textit{et al.} \cite{Jejjala1}, for our present purposes the important flow-down consequence is that one obtains a closed form Friedmann equation
\be
E^2(z)=\Omega_m(1+z)^3+(1-\Omega_m)\frac{1-b(\xi)}{1-b(\xi_0)}\,,
\ee
where
\be
b(\xi)=\left(1+\xi+\frac{1}{2}\xi^2+\frac{1}{6}\xi^3\right)e^{-\xi}\,,
\ee
\be
\xi=\frac{\xi_0}{1+z}\;.
\ee
We thus have again a parametric extension of $\Lambda$CDM, where the only non-standard parameter is the dimensionless scaling parameter $\xi_0$. In this case the $\Lambda$CDM limit is recovered in the limit $\xi_0\to\infty$. Figure \ref{fig3} identifies, as a function of the matter density and the scaling parameter, the redshift at which the two terms in the Friedmann equation have identical values. This shows that for values $\xi_0\ge 10$ the model is indistinguishable from $\Lambda$CDM, and Occam's razor (or analogous statistical model selection arguments) would therefore indicate that the additional parameter is unwarranted. A recent work \cite{Jejjala2} has a simplistic comparison with Hubble parameter data, and claims a preference for a finite value of $\xi_0$. They also suggest that in this model one obtains a phantom (and therefore physically problematic) dark energy equation of state. Be that as it may, in what follows we provide a more robust analysis, also relying on an extended data set.

\begin{figure}
\begin{center}
\includegraphics[width=\columnwidth]{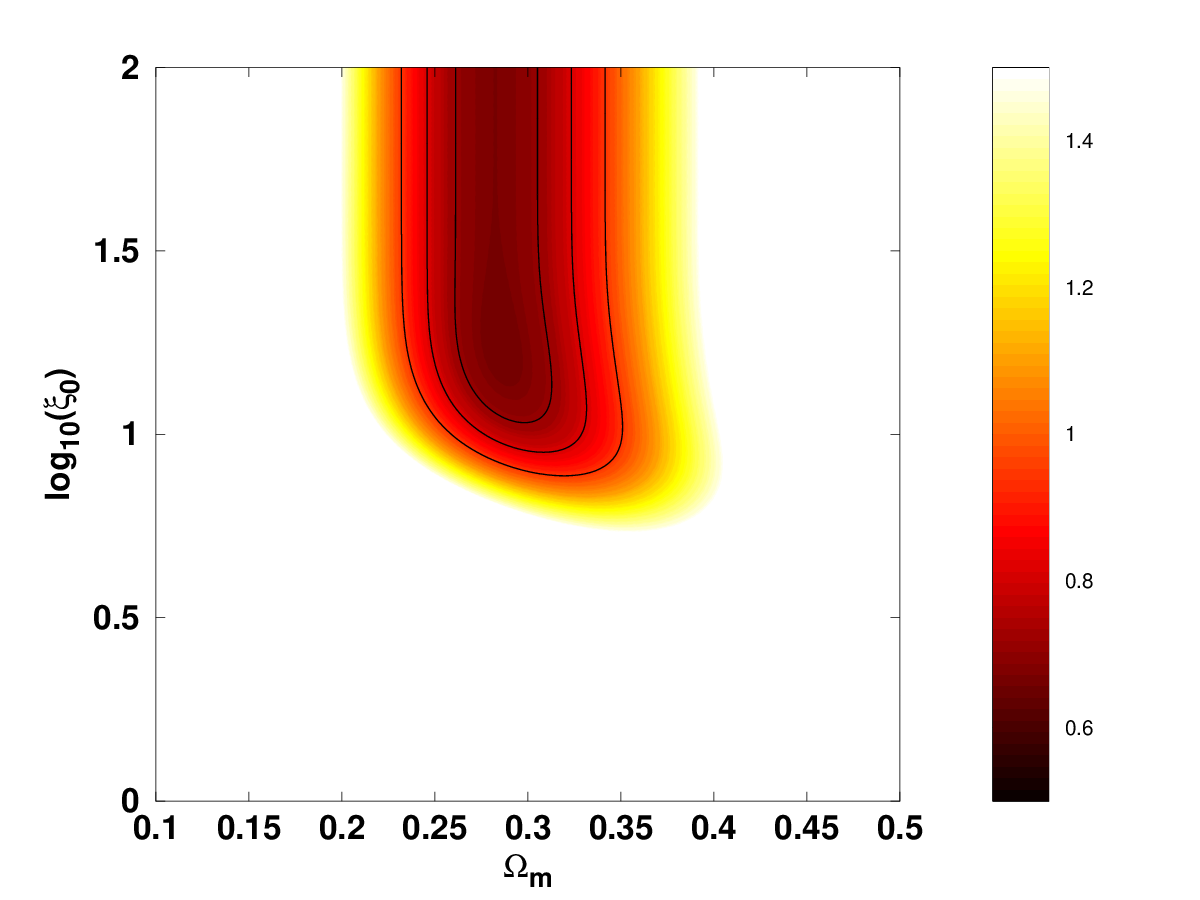}
\includegraphics[width=\columnwidth]{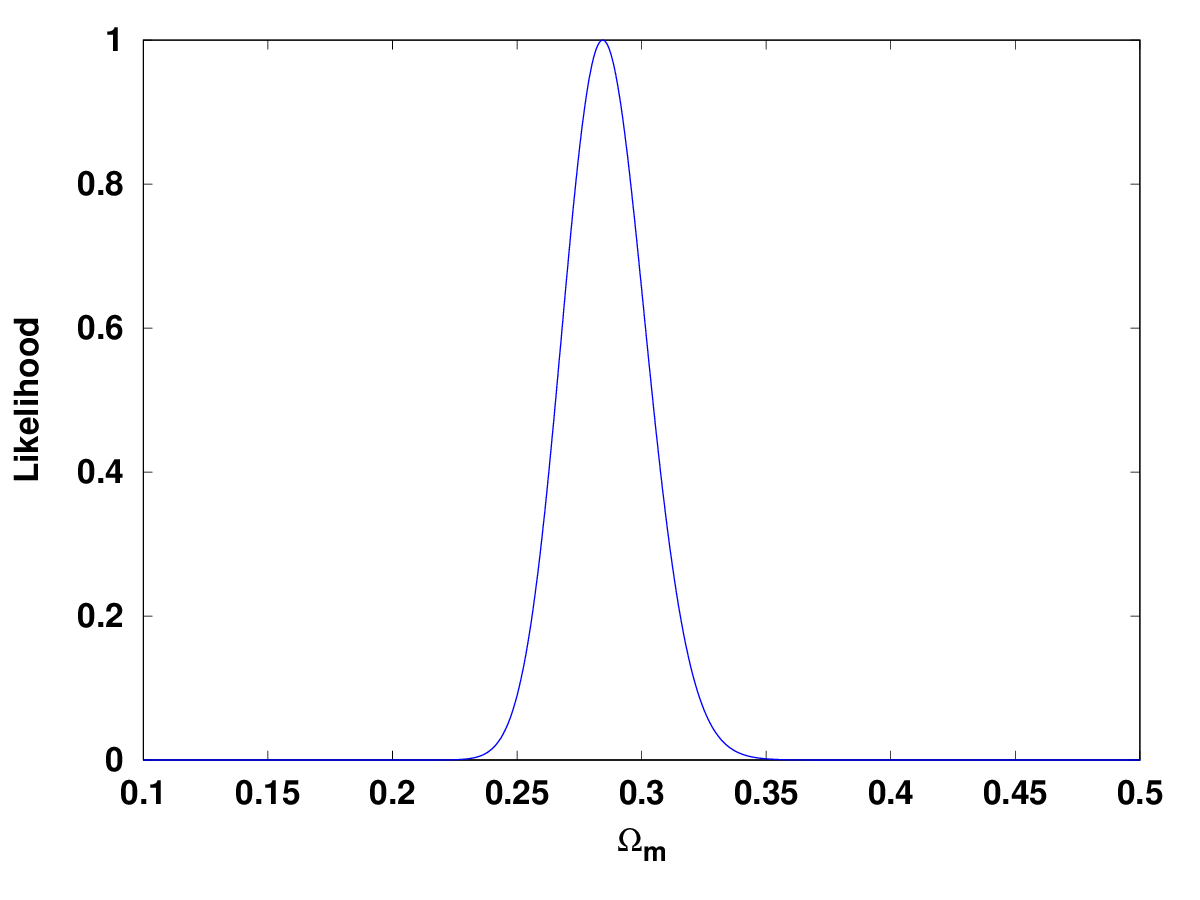}
\includegraphics[width=\columnwidth]{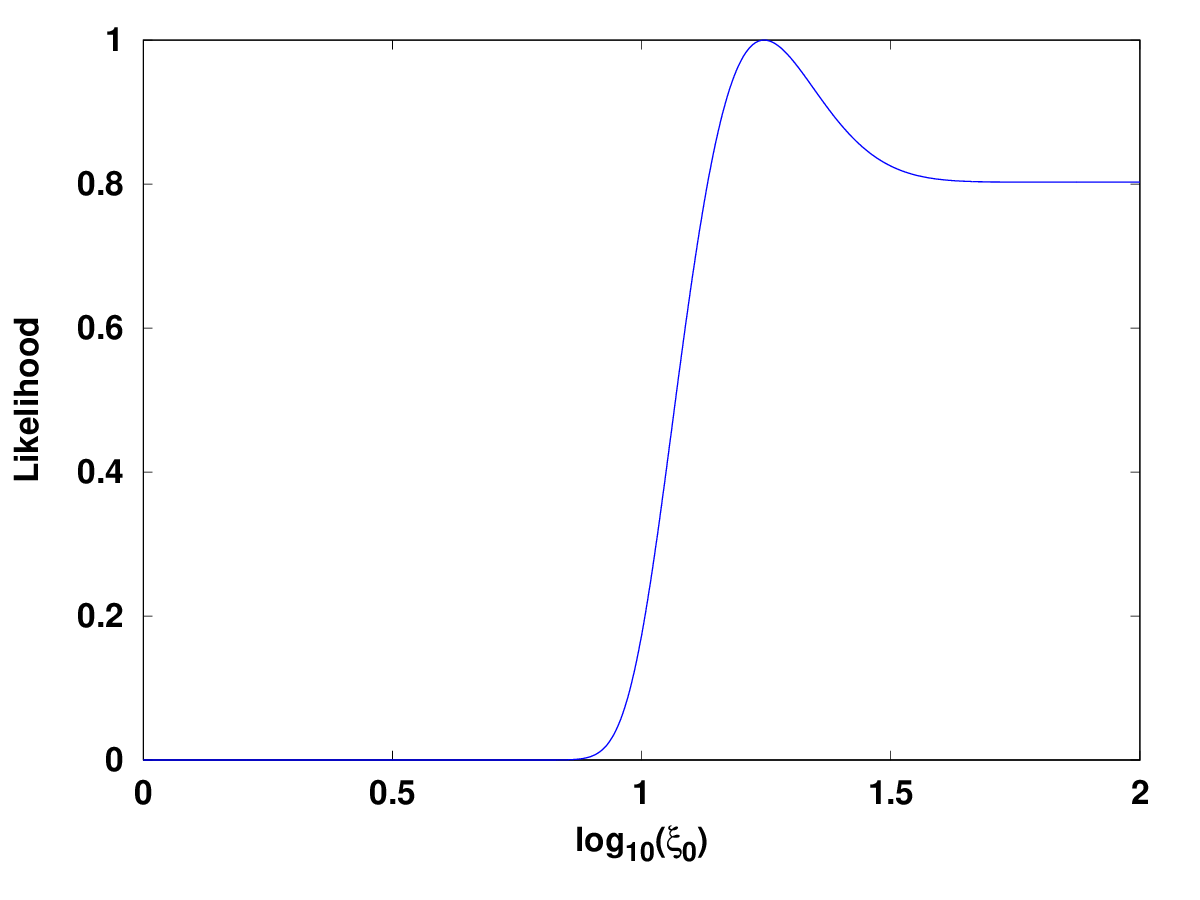}
\caption{Constraints on the infinite statistics model, under the assumptions specified in the main text. In the top panel the black lines shows the one, two and three sigma confidence levels, while the color map depicts the reduced chi-square. The middle and bottom panels show the one-dimensional posterior likelihoods for each parameter.}\label{fig4}
\end{center}
\end{figure}

The results of this analysis are shown in Figure \ref{fig4}. For the matter density we use the same uniform prior as in the previous section, $\Omega_m\in[0.1,0.5]$, while for the scaling parameter we use a uniform prior in its logarithm, $\log_{10}(\xi_0)\in[0,2]$. As in the model in the previous section there is no degeneracy between the two parameters, and we have explicitly verified that equivalent results would be obtained from different choices of priors (e.g., a uniform prior in $\xi_0$). We do not fully reproduce the results of \cite{Jejjala2}, and instead find no statistically significant preference for a finite value of $\xi_0$. Specifically our one-sigma constraint on the matter density is
\begin{equation}
    \Omega_{m}= 0.28 \pm 0.02\,,
\end{equation}
while for the logarithm of the dimensionless scaling parameter we have the two-sigma lower limit
\begin{equation}
    {\textrm log}_{10}(\xi_0) > 0.99\,.
\end{equation}
As in the previous case, we find consistency with the $\Lambda$CDM model.

\section{The Regge-Teitelboim model}

In Regge-Teitelboim gravity, one embeds the space–time manifold in a fixed higher dimensional background, and further assumes that the embedding coordinates, rather than the metric tensor, are the dynamical degrees of freedom \cite{Regge}. (We note that this approach has been criticized by other authors \cite{Deser}.) In principle one can choose different embeddings, and these choices will lead to correspondingly different cosmological solutions. A common feature of these cosmological solutions is the presence of additional source terms in the standard Einstein equations that do not come from the energy-momentum tensor. One may therefore consider the possibility that these additional terms could be responsible for the recent acceleration of the universe. Here we study three previously considered such scenarios \cite{Davidson,Fabi,Stern}, which stem from different five-dimensional embedding choices. Note that in these models the vacuum energy density (or cosmological constant) is assumed to vanish, so the models do not have a $\Lambda$CDM limit.

\subsection{The Davidson model}

In the model first considered by Davidson \cite{Davidson}, an embedding in a flat five-dimensional background leads to a dimensionless Friedmann equation which can be written
\be
E^2(z)-\Omega_m(1+z)^3-\Omega_k(1+z)^2=\frac{\Omega_\mu (1+z)^4}{\sqrt{E^2-\Omega_k(1+z)^2}}\,,
\ee
where $\Omega_k=-k/(a_0H_0)^2$ is the usual spatial curvature density parameter and we have defined
\be
\Omega_\mu=\frac{\mu}{27H_0^3a_0^4}\,,
\ee
and $\mu$ is an integration constant, coming from the choice of the embedding \cite{Davidson}, which in practical terms describes the deviation from the standard Friedmann equation (without a cosmological constant). This is subject to the consistency condition
\be\label{consistencyD}
1-\Omega_m-\Omega_k=\frac{\Omega_\mu}{\sqrt{1-\Omega_k}}\,.
\ee

Alternatively one may note, as shown in \cite{Davidson} that the effective equation of state of the additional term is $w_{\rm eff}=p_{\rm eff}/\rho_{\rm eff}=-1/9$, so we can also write
\be
E^2(z)=\Omega_m(1+z)^3+\Omega_k(1+z)^2+(1-\Omega_m-\Omega_k)(1+z)^{8/3}\,,
\ee
suggesting that this will not provide a good fit to the data.

\begin{figure}
\begin{center}
\includegraphics[width=\columnwidth]{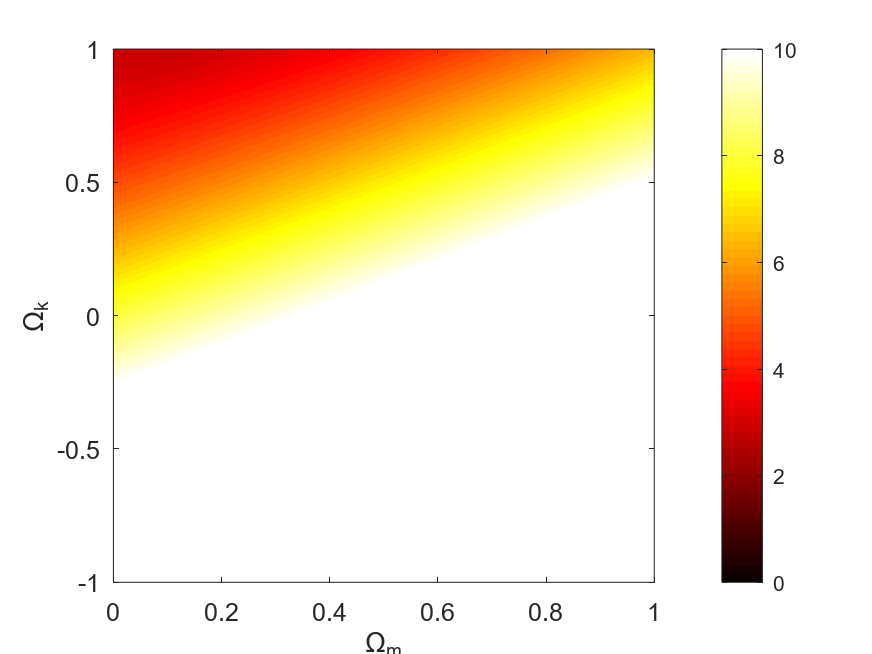}
\includegraphics[width=\columnwidth]{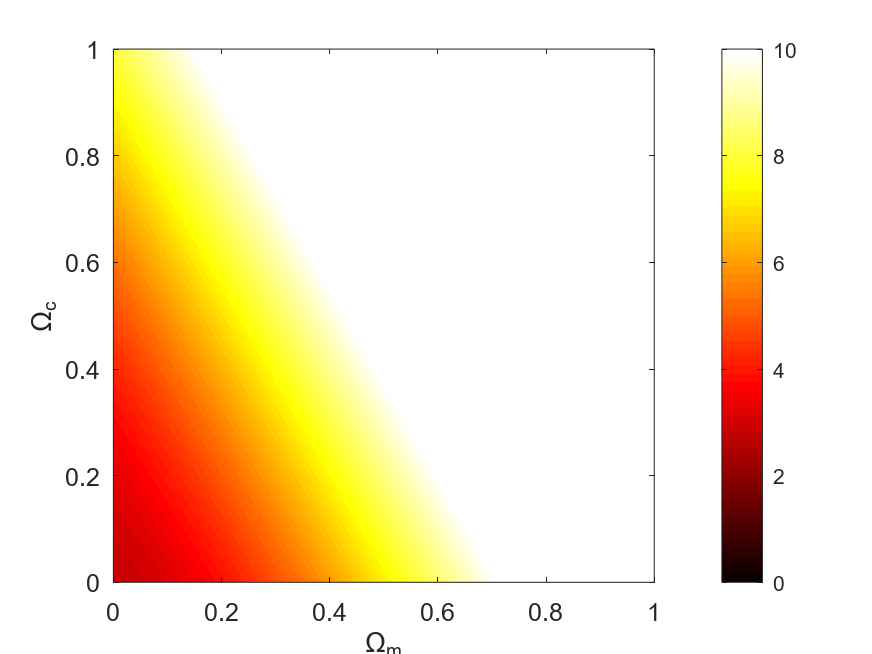}
\caption{The reduced chi-square, indicated by the color map, for the (possibly) physically realistic regions of parameter space in the models of Davidson (top panel) and Fabi \textit{et al} (bottom panel). Notice that the color map spans the range from 0 to 10; in both cases the minimum value of the reduced chi-square is $\chi^2_\nu\sim2.9$.}\label{fig5}
\end{center}
\end{figure}

The result of our likelihood analysis, for the broad range of values of $\Omega_m\in[0.0,1.0]$ and $\Omega_k\in[-1.0,1.0]$, is shown in the top panel of Figure \ref{fig5}: clearly the model does not fit the data, and the best reduced chi-square in this parameter range is around $\chi^2_\nu\sim2.9$; for comparison, for the standard $\Lambda$CDM compared to the datasets which we are considering in this work the reduced chi-square is $\chi^2_\nu\sim2.9$. We explicitly checked that the two above parameterizations lead to similar results. It is curious to note that the best (or, more rigorously, least bad) fit would correspond to an empty universe with $k=-1$, which has some similarities with another model in the same class which we discuss in what follows, but in any case this particular model does not fit the data.

\subsection{The Fabi \textit{et al.} model}

In the model of Fabi \textit{et al.} \cite{Fabi}, a different embedding in a flat five-dimensional background leads to the following Friedmann equation
\be
H^2+\frac{k}{a^2}=\frac{8\pi G}{3}\left[\rho+\frac{c_0}{a^3\sqrt{1+k{\dot a}^2}}\right]\,,
\ee
where $c_0$ is again an integration constant. In this case the Friedmann equation is only changed for $k\neq0$. For $k=0$ one gets the standard Friedmann equation but with a rescaled matter density (or, equivalently, a rescaled Newtons's constant), which clearly can't fit the data in the absence of a cosmological constant. Fabi \textit{et al.} claim that $k=-1$ leads to low-redshift acceleration. More accurately, we find that this is a necessary (but not sufficient) condition for acceleration.

Defining, for convenience,
\be
\Omega_c=\frac{8\pi Gc_0}{3H_0^2a_0^3}\,,
\ee
we can re-write the Friedmann equation as
\be
E^2(z)+\frac{k(1+z)^2}{(a_0H_0)^2}=\Omega_m(1+z)^3+\frac{\Omega_c(1+z)^3}{\sqrt{1+\frac{kE^2(z)(a_0H_0)^2}{(1+z)^2}}}\,.
\ee
For each choice of the curvature parameter $k$, the normalization condition $E(0)=1$ enables the determination of the $(a_0H_0)^2$ parameter, which must obviously be positive, through
\be
1+\frac{k}{(a_0H_0)^2}=\Omega_m+\frac{\Omega_c}{\sqrt{1+k(a_0H_0)^2}}\,.
\ee
Interestingly, \cite{Fabi} also show that the effective equation of state of the additional term is $w_{\rm eff}=p_{\rm eff}/\rho_{\rm eff}=-1/9$, just like in the Davidson case. It follows that, for the most relevant $k=-1$ case, we can also write the dimensionless Friedmann equation as
\be
E^2(z)=\Omega_m(1+z)^3+\Omega_c(1+z)^{8/3}+(1-\Omega_m-\Omega_c)(1+z)^2\,;
\ee
despite the different free parameters, its similarity with the one for the Davidson case is manifest. The corresponding likelihood analysis can be seen in the bottom panel of Figure \ref{fig5}, and naturally the result is the same as in the Davidson case.

\subsection{The Stern \& Xu model}

The assumption of a flat five-dimensional embedding, which underlies the models in the two previous sub-sections, has recently been relaxed by Stern \& Xu \cite{Stern}, who have briefly explored non-flat cases. Specifically, they claim that an embedding in five-dimensional de Sitter space can lead to late-time accelerating solutions.

\begin{figure}
\begin{center}
\includegraphics[width=\columnwidth]{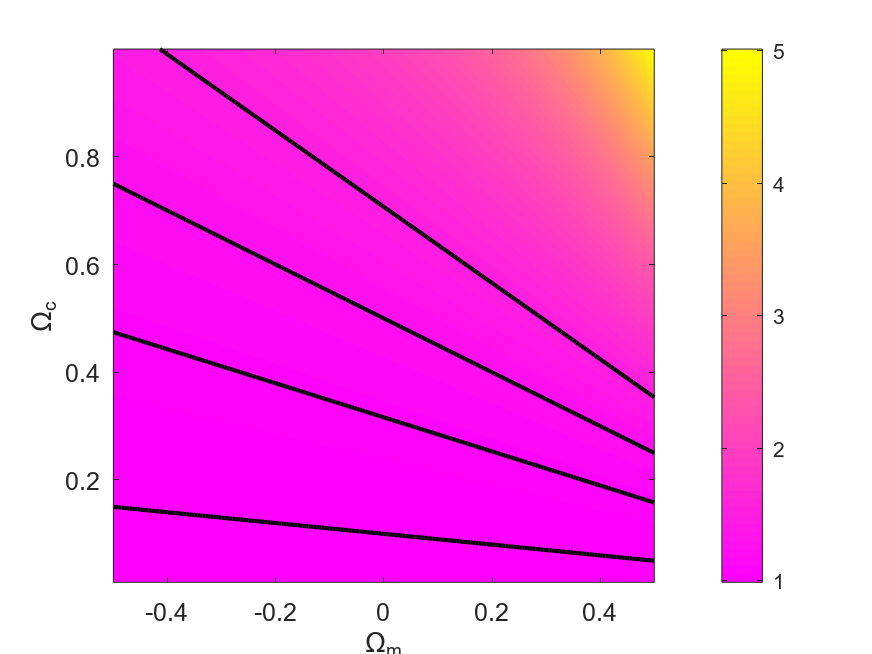}
\caption{The value of the dimensionless parameter $(LH_0)^2$, as a function of the matter density $\Omega_m$ and the additional model parameter $\Omega_c$, depicted by the color map. The black solid lines identify the locus of values of this parameter of 1.01, 1.1, 1.25, and 1.5, respectively from the bottom to the top of the plot.}\label{fig6}
\end{center}
\end{figure}
\begin{figure*}
\begin{center}
\includegraphics[width=\columnwidth]{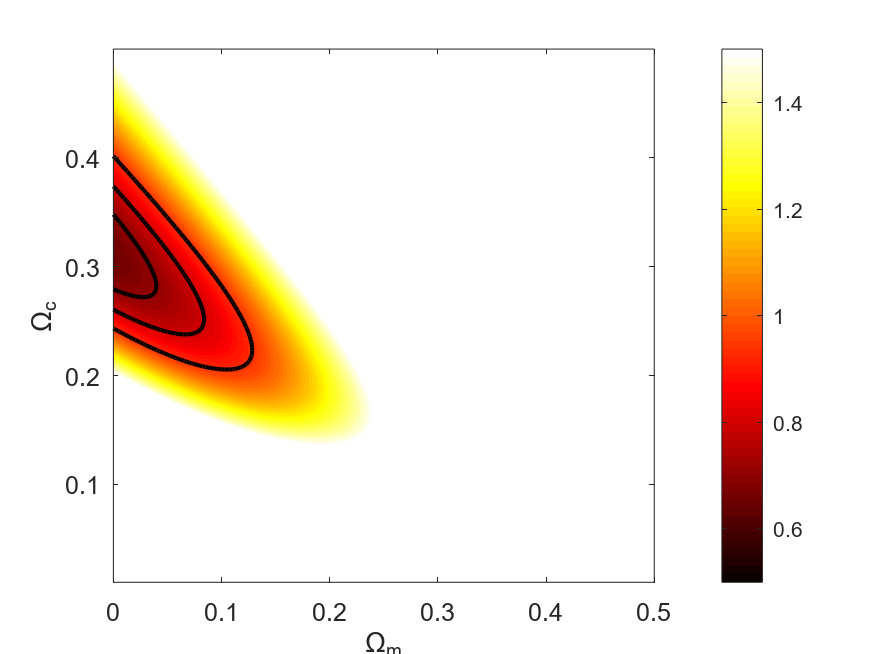}
\includegraphics[width=\columnwidth]{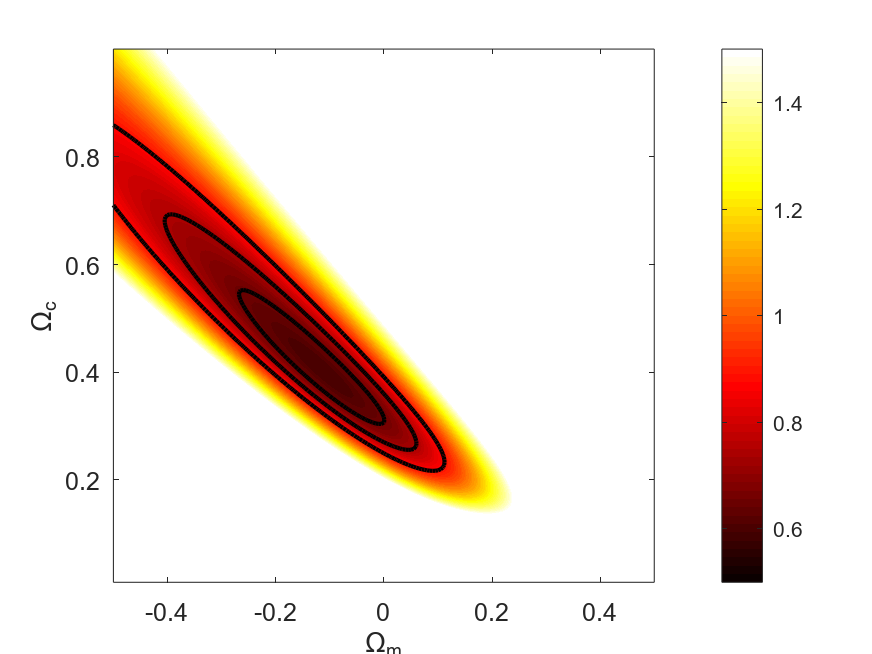}
\includegraphics[width=\columnwidth]{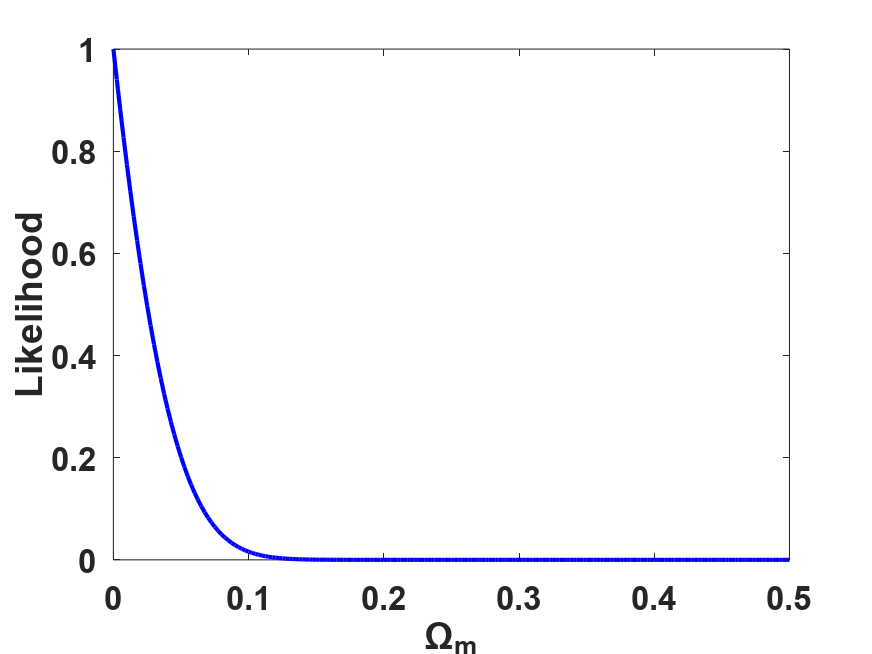}
\includegraphics[width=\columnwidth]{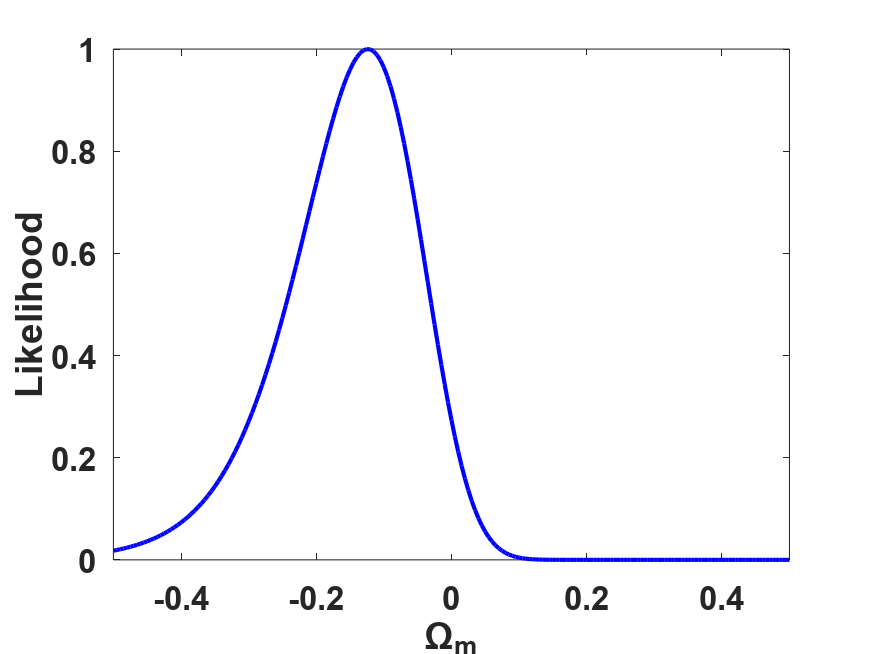}
\includegraphics[width=\columnwidth]{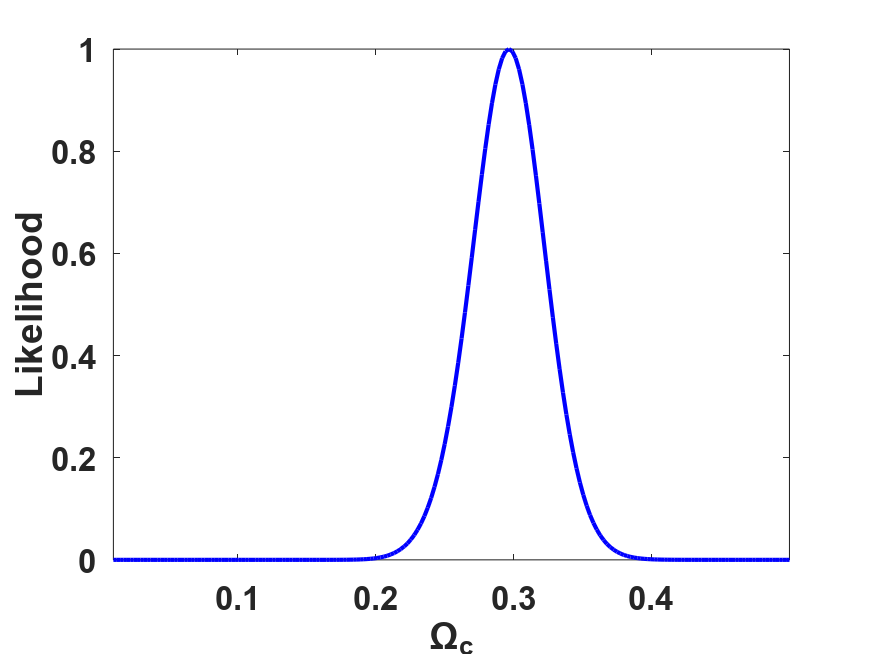}
\includegraphics[width=\columnwidth]{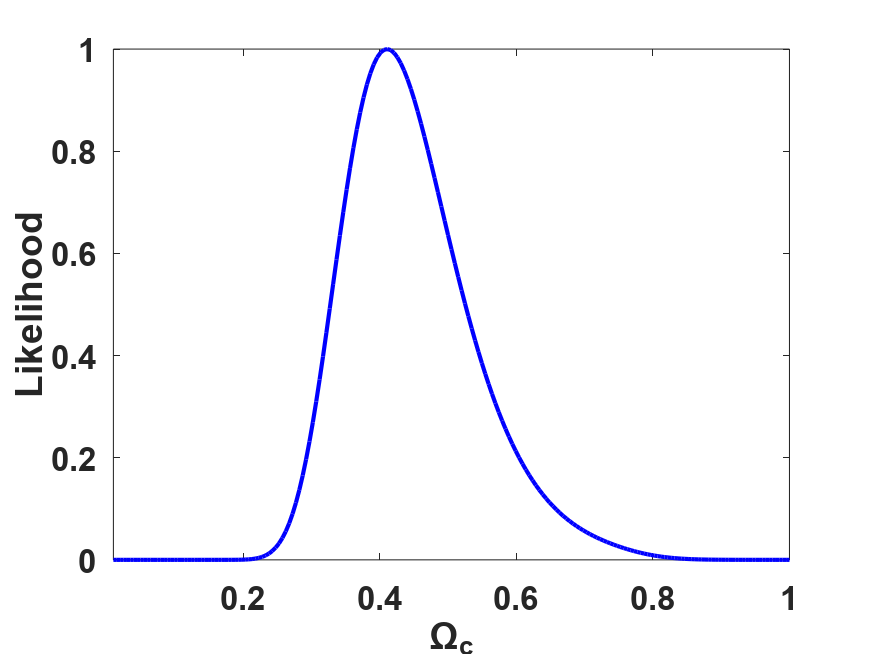}
\caption{Constraints on the Stern \& Xu realization of the Regge-Teitelboim model. In the top panels the black lines shows the one, two and three sigma confidence levels, while the color map depicts the reduced chi-square. The middle and bottom panels show the one-dimensional posterior likelihoods for each parameter. Left-side panels show the results with a $\Omega_m\ge0$ prior, while in right-side panels this assumption is relaxed.}\label{fig7}
\end{center}
\end{figure*}

Defining $L$ to be the (de Sitter) curvature radius, one can now write the dimensionless Friedmann equation as
\be
E^2(z)=\Omega_m(1+z)^3+\frac{\Omega_c(1+z)^4}{\sqrt{(LH_0)^2E^2-1}}\,,
\ee
or equivalently
\be
E^2(z)=\Omega_m(1+z)^3+(1-\Omega_m)\frac{\sqrt{(LH_0)^2-1}}{\sqrt{(LH_0)^2E^2-1}}(1+z)^4\,,
\ee
with the additional relation between the model parameters
\be
(LH_0)^2=1+\left(\frac{\Omega_c}{1-\Omega_m}\right)^2\,.
\ee
Note that this parameter space is further constrained, since there are combinations of parameters that would lead to an unphysical behaviour of $E(z)$, especially at higher redshifts. This effectively imposes a theoretical prior on this model's parameter space. In the analysis which we report this prior has been imposed, and no unphysical behaviour occurs for the ranges of the model parameters depicted in the figures resulting from the likelihood analysis which follow, and for the redshift range relevant for the data which we are considering. Nevertheless, and as a caveat for the reader, we note that we are proceeding on the phenomenological assumption that this is a low-redshift approximation to the behaviour of a physically more robust underlying mechanism.

Figure \ref{fig6} depicts the value of this parameter, as a function of $\Omega_m$ and $\Omega_c$, for a wide range of values of the two parameters. A previous qualitative analysis \cite{Stern} suggests that this model can fit low-redshift Hubble parameter measurements, even in the absence of matter. We note that the data set used in \cite{Stern} is a subset of the Farooq \textit{et al} dataset, while in the present analysis we use the full Farooq \textit{et al} dataset together with Type Ia supernovae, as has been previously discussed. In what follows we present a more robust statistical analysis.

Figure \ref{fig7} summarizes our constraints on this model. We start by assuming the uniform priors $\Omega_m\in[0.0,0.5]$ and $\Omega_c\in[0.0,0.5]$, for which the results are shown in the left-hand side panels of the figure. We confirm that the model fits the data, and indeed it overfits it, with the reduced chi-square at the best fit being $\chi^2_\nu\sim0.7$. (For comparison, fitting the standard CPL phenomenological dark energy model to the same data yields a reduced chi-square at the best fit of $\chi^2_\nu\sim0.9$ \cite{Fernandes}.) We derive one-sigma posterior constraint
\be
\Omega_c=0.30\pm0.03\,,
\ee
and a two-sigma upper limit
\be
\Omega_m<0.06\,,
\ee
which together lead to a dimensionless curvature radius $LH_0\sim1.04$. Looking at the shape of the likelihood, one may infer that the peak of the statistical likelihood is elsewhere (outside the chosen prior range), which can easily be confirmed by relaxing the previous assumption and allowing (again, purely phenomenologically) for negative values of $\Omega_m$. The results in this case are shown in the right-hand side panels of Figure \ref{fig7}. In this case the reduced chi-square at the best fit decreases to $\chi^2_\nu\sim0.6$, and we obtain the one-sigma posterior constraint
\be
\Omega_c=0.41_{-0.08}^{+0.10}\,
\ee
\be
\Omega_m=-0.13_{-0.10}^{+0.09}\,.
\ee
As expected the two parameters are negatively correlated, and in this wider parameter space the preferred value of $\Omega_c$ increases, while that of the matter density is correspondingly decreased. Naturally a negative matter density is observationally questionable, although it is only a one sigma-result and, as has been pointed out, the model quite overfits the data. The preferred dimensionless curvature radius is comparatively less impacted, now having a value $LH_0\sim1.06$.

As a final and again purely phenomenological exercise, we can also consider the case where a cosmological constant is included in the model, in addition to the higher-dimensional term. In other words, the Friedmann equation becomes 
\be
E^2(z)=\Omega_m(1+z)^3+\Omega_\Lambda+\frac{\Omega_c(1+z)^4}{\sqrt{(LH_0)^2E^2-1}}\,,
\ee
with
\be
(LH_0)^2=1+\left(\frac{\Omega_c}{1-\Omega_m-\Omega_\Lambda}\right)^2\,.
\ee
In this case this is also a parametric extension of $\Lambda$CDM, but it is still the case that this parameter space is further constrained, and some combinations of parameters would lead to an unphysical behaviour of $E(z)$. Admittedly, from a statistical perspective this is a moot exercise because the simpler model already overfits the data. On the other hand, from a phenomenological point of view it is a useful exercise to infer, despite the expected degeneracies between the parameters, the extent to which the data selects between the various terms (and their redshift dependencies).

\begin{figure*}
\begin{center}
\includegraphics[width=\columnwidth]{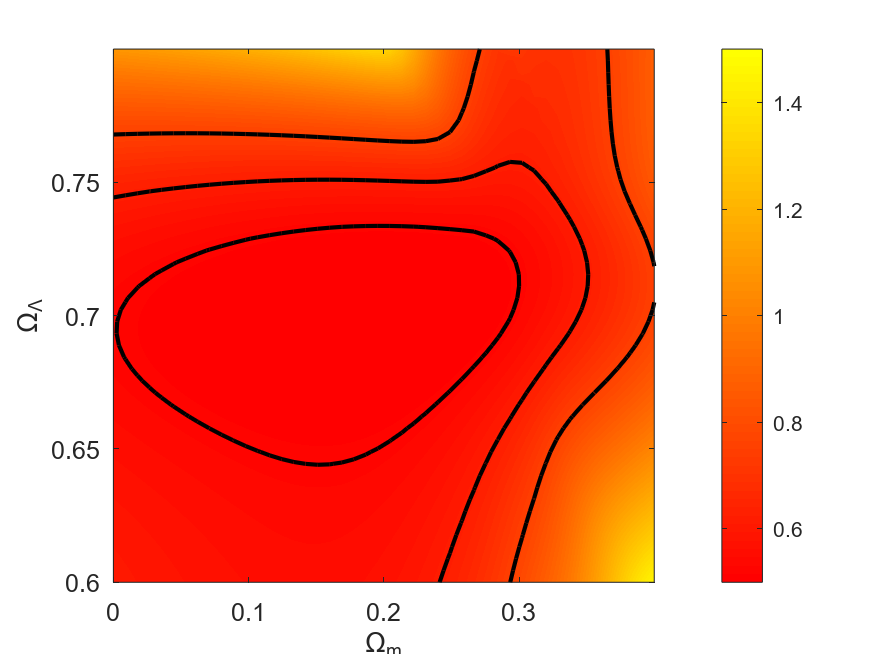}
\includegraphics[width=\columnwidth]{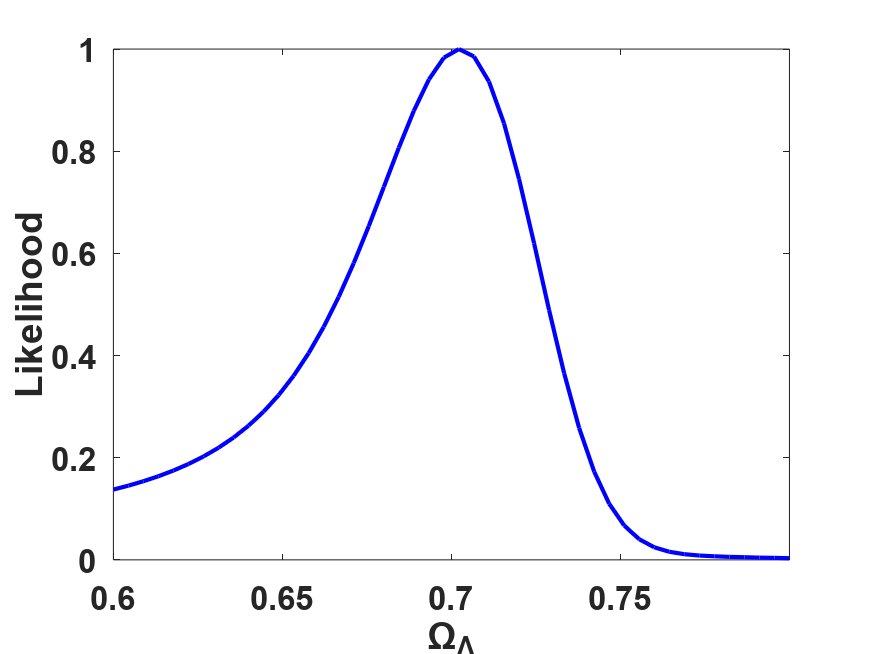}
\includegraphics[width=\columnwidth]{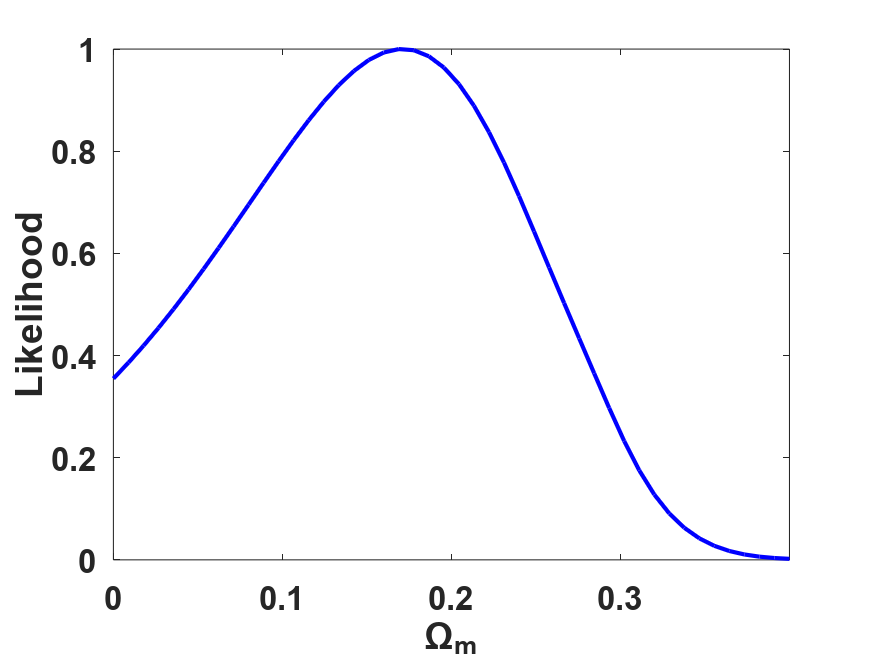}
\includegraphics[width=\columnwidth]{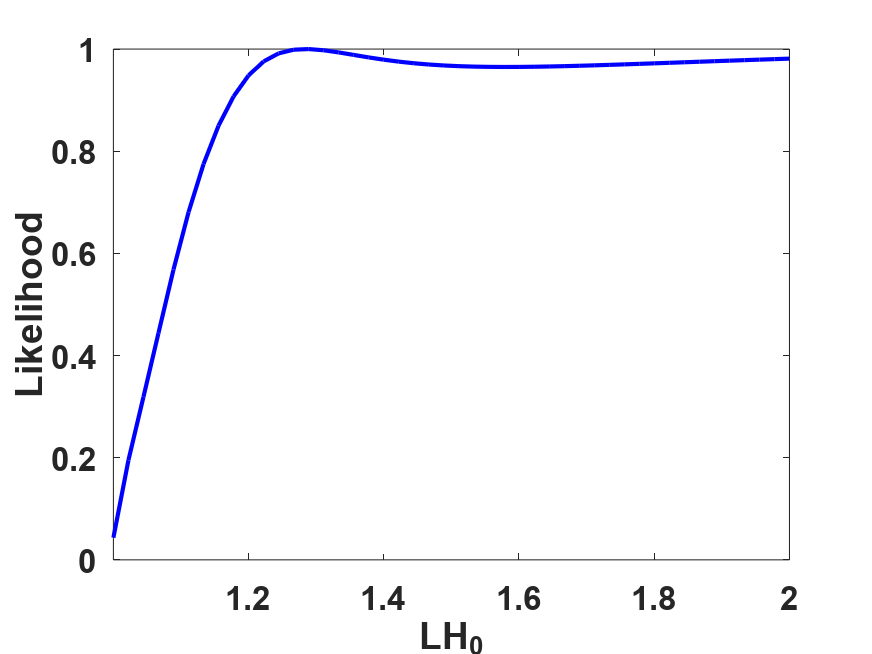}
\caption{Constraints on the Stern \& Xu realization of the Regge-Teitelboim model, allowing for the presence of a cosmological constant. In the top left panel the black lines shows the one, two and three sigma confidence levels, while the color map depicts the reduced chi-square. The other panels show the one-dimensional posterior likelihoods for each parameter.}\label{fig8}
\end{center}
\end{figure*}

Figure \ref{fig8} shows the results of our analysis exploring the parameter space in the neighbourhood of the standard flat $\Lambda$CDM model parameters and using the set $(\Omega_m, \Omega_\Lambda, LH_0)$ as independent parameters, with uniform (uninformative) priors on each of them. Interestingly, we find that the statistically preferred value of the cosmological constant is still the standard one, while that of the matter density is decreases with respect to the standard one, being partially offset by the new term in the Friedmann equation. Specifically, we find the one-sigma constraints
\be
\Omega_\Lambda=0.70_{-0.03}^{+0.02}
\ee
\be
\Omega_m=0.17_{-0.11}^{+0.08}\,,
\ee
together with the two-sigma lower limit
\be
(LH_0)>1.02\,.
\ee

For comparison, we have also considered the alternative choice of using the set $(\Omega_m, \Omega_\Lambda, \Omega_c)$ as independent parameters, again with uniform priors on each of them. This leads to results which, within their uncertainties, are consistent with the above ones. Specifically, we now have the one-sigma constraints
\be
\Omega_\Lambda=0.71_{-0.02}^{+0.03}
\ee
\be
\Omega_m=0.18_{-0.10}^{+0.07}\,,
\ee
and a two-sigma lower limit in the additional parameter
\be
\Omega_c>0.02\,.
\ee
As expected, the reduced chi-square at the best fit is still $\chi^2_\nu\sim0.6$ in both cases.

\section{Conclusions}
\label{concl}

We have compared three classes of modified gravity models for the low-redshift acceleration of the universe against low-redshift background cosmological observations. Each of them stems from different (and possibly questionable) physical assumptions, but our analysis has been a purely phenomenological one, addressing the question of the extent to which the models are in agreement with available data and, more specifically of how this data constrains model parameters.

For the fist two models classes---the Lifshitz model recently explored by Berechya \& Leonhardt \cite{Berechya} and the infinite statistics model introduced by Jejjala \textit{et al.} \cite{Jejjala1,Jejjala2}---the outcome is effectively the same. At least for the low-redshift range under consideration, they are one-parameter extensions of $\Lambda$CDM, and the data constrains this parameter to be consistent with and very close to its 'null' value. In other words, there is no evidence of deviations from $\Lambda$CDM. Moreover, for the Lifshitz model, this additional parameter (a dimensionless coupling parameter) has a theoretically predicted value, or at least one that would be preferred by the physical assumptions underlying the model. Our analysis demonstrates that this preferred value of the coupling is ruled out at more than six standard deviations.

For Regge-Teitelboim gravity, the situation is different. Here there is no cosmological constant, and therefore they are not parametric extensions of $\Lambda$CDM. Instead, the recent acceleration of the universe must be due to a different physical mechanism. Specifically, this is an additional source term in the standard Einstein equations, which does not come from the usual four-dimensional energy-momentum tensor but emerges from the higher-dimensional nature of the model in which the standard space-time manifold is embedded. In this class of models one has some freedom in the choice of embeddings, and we have provided constraints on three such choices, all previously considered in the literature. Two of these choices are thoroughly ruled out, as they would have an effective dark energy equation of state $w_{\rm eff}=-1/9$. The third choice, recently considered by Stern \& Xu \cite{Stern} is compatible with the data we have considered, and actually overfits it.

It is interesting to consider why this model requires a lower than standard matter density to provide a good fit. Indeed, in the form presented by Stern \& Xu the statistically preferred value would be a negative matter density, though naturally this is purely a point of statistics and not a point of physics. One does know that the universe contains baryons; whether or not it contains dark matter is may be more debatable. As can be seen in the case of  Fabi \textit{et al.}, and even in the original Regge-Teitelboim work, a rescaled matter density seems to be a common feature in these models. This interpretation is commensurate with the fact that, if one phenomenologically extends these models by allowing for the possibility of a cosmological constant, then the low-redshift data used in our work still prefers the standard value of the cosmological constant, $\Omega_\Lambda\sim0.7$, but a lower value of the matter density, $\Omega_m\sim0.2$. That said, we hasten to add that these models can, at best, be low-redshift approximations to some more fundamental models since for a significant range of the model's parameter space one would find an unphysical behaviour of $E(z)$, at least at high redshifts.

In any case, our analysis does highlight the tight constraints on the allowed low-redshift deviations from the standard $\Lambda$CDM background evolution. Naturally, more stringent constraints can be obtained by extending our analysis beyond low-redshift background cosmology data, e.g. by including cosmic microwave background data. While this may be moot for most of the models we have considered (because they are already ruled out, our constrained by low-redshift data to be effectively indistinguishable from $\Lambda$CDM), it would be interesting to carry out this analysis for Stern \& Xu model, or physically motivated extensions thereof.

\acknowledgments
This work was financed by Portuguese funds through FCT - Funda\c c\~ao para a Ci\^encia e a Tecnologia in the framework of the project 2022.04048.PTDC.  CJM also acknowledges FCT and POCH/FSE (EC) support through Investigador FCT Contract 2021.01214.CEECIND/CP1658/CT0001. 

\bibliography{article}
\end{document}